\documentclass[12pt]{amsart}
\usepackage{amssymb}
\usepackage{amsmath}
\usepackage{amsfonts}
\usepackage{mathrsfs}
\usepackage{graphicx}
\usepackage{color}
\usepackage[onehalfspacing]{setspace}
\usepackage{caption}
\usepackage{enumerate}
\usepackage[round]{natbib}
\usepackage{appendix}
\usepackage{lscape}
\usepackage{subcaption}
\usepackage{graphicx}
\usepackage{amsfonts}
\usepackage{placeins}
\usepackage[utf8]{inputenc}
\usepackage{charter}
\usepackage[colorlinks=true,citecolor=blue,urlcolor=blue,pdfpagemode=UseNone,pdfstartview=FitH]{hyperref}
\usepackage{apptools}
\usepackage{multibib}
\usepackage{multirow}
\usepackage{booktabs}
\newcites{main,supp}{References,References}
\makeatletter
\def\section{\@startsection{section}{1}
	\z@{1.0\linespacing\@plus\linespacing}{.8\linespacing}{\Large}}

\def\subsection{\@startsection{subsection}{2}
	\z@{.8\linespacing\@plus.7\linespacing}{.7\linespacing}{\large}}

\def\subsubsection{\@startsection{subsubsection}{3}
	\z@{.5\linespacing\@plus.7\linespacing}{-.5em}{\normalfont\bfseries}}
\makeatother

\numberwithin{equation}{section}

\theoremstyle{definition}
\newtheorem{definition}{Definition}[section]

\theoremstyle{definition}

\theoremstyle{definition}

\title{}
\begin{document}
	\vspace*{3ex minus 1ex}
	\begin{center}
		\Large \textsc{Quantifying Theory in Politics: \\ Identification, Interpretation and the Role of Structural Methods}
		\bigskip
	\end{center}
	
	\date{%
		\today%
	}

\vspace*{2ex minus 1ex}
	\begin{center}
		Nathan Canen and Kristopher Ramsay\\
		\medskip
		\medskip
	\end{center}
	
	\thanks{\textbf{Canen:} University of Houston and Research Economist at NBER.\\ \textbf{Ramsay:} Princeton University. \\ We would like to thank Scott Ashworth, Andy Eggers, Sean Gailmard, Mike Gibilisco, Bobby Gulotty,  Gleason Judd, Amanda Kennard, Korhan Kocak, Monika Nalepa, V\'{i}tor Possebom, Joseph Ruggiero, Kevin Song, Tara Slough, Dustin Tingley, Scott Tyson, participants at the University of Chicago Workshop on Uses of Formal Theory for comments. The ideas in this paper have been improved and shaped by numerous conversations over many years in various venues and meetings. } 

\begin{abstract}
    
   The best empirical research in political science clearly defines substantive parameters of interest, presents a set of assumptions that guarantee its identification, and uses an appropriate estimator. We argue for the importance of explicitly integrating rigorous theory into this process and focus on the advantages of doing so. By integrating theoretical structure into one's empirical strategy, researchers can quantify the effects of competing mechanisms, consider the ex-ante effects of new policies, extrapolate findings to new environments, estimate model-specific theoretical parameters, evaluate the fit of a theoretical model, and test competing models that aim to explain the same phenomena. As a guide to such a methodology, we provide an overview of structural estimation, including formal definitions, implementation suggestions, examples, and comparisons to other methods.  
\end{abstract}
\maketitle
\medskip
		
		{\noindent \textsc{Keywords.} Quantitative Methods, Formal Theory, Identification, Structural Methods, Counterfactuals, Research Designs.}
		\medskip
	
	\newpage
\section{Introduction}


All quantitative empirical analysis is based on models. Counterfactuals are derived from models, as is the identification of parameters and interpretation of their estimates. Indeed, it is theoretical assumptions that bridge the gap between estimates and their interpretation \citep{lundberg2021whata}. The validity of such assumptions, and the choice of which specification to use, is necessarily a result of the connection between the empirical strategy and underlying theory, whether motivated formally or qualitatively. Thus, it is critical to think carefully about the underlying model.

We provide a framework for empirical analysis grounded in this observation. Since all empirical quantities of interest are defined relative to some theoretical model, the framework can be used by any researcher. It does not depend on the use of `structural' or `reduced-form' methods.

Our framework begins with defining an empirical target, which will depend on the research question and theoretical interest. Then, the researcher specifies their underlying modeling assumptions and specification. Together, these allow the researcher to interpret their findings and can guarantee statistical identification for their parameter(s) of interest. Finally, the researcher proposes an estimator to estimate the desired parameter from a sample.

To see why our framework is useful, consider the following example.  Suppose a researcher has data on vote shares for party $p$ in district $d$ and on a district-level characteristic, $X_d$ (e.g., average education levels, share of women in the district, etc.), and decides to run the following linear regression: 
\begin{equation}
\log(vote~share_{p,d}) = \alpha + \beta X_d + \varepsilon_{p,d}.\label{example_intro}
\end{equation}
When should this researcher interpret $\beta$ as: (i) the slope of the linear best fit between the outcome and $X_d$, (ii) a (semi)-elasticity of $X_d$ on vote shares of $p$, (iii) the Average Treatment Effect from a policy which increased $X_d$, (iv) preferences from voters with characteristic $X_d$ for party $p$, (iv) the effect from a counterfactual policy that would increase $X_d$? All five interpretations appear in the literature when discussing the results of this regression \citep{bartels2001presidential,gerber1998estimating, nadeau2001national, hansford2010estimating}, but the quantity in each case is defined relative to a different underlying model. Furthermore, these theoretical quantities cannot be differentiated from the estimates of (\ref{example_intro}). And yet, the choice between (i)-(v) can be very consequential: each interpretation may yield vastly different theoretical, policy and welfare conclusions. In this way, the analysis is reliant on a theoretical model all the way down.

Motivated by this, in the first part of this article, we elaborate on a framework for analysis that pays careful attention to their model. This framework simply recognizes that all parameters of interested are defined relative to some model. That is, there is no atheoretical, or model free statistics. Quantitative analysis is models all the way down. The validity or usefulness of assumptions that allow statistical analysis, and the choice of which specification to use, is inherently due to theory. 


In the second part of the paper, starting with Section \ref{section_structural}, we discuss empirical methods for researchers who wish to identify and estimate parameters that are defined relative to a formal theoretical model of politics. Examples include preferences (e.g., ideologies), measures of welfare, or parameters governing agent behavior (e.g., the magnitude of strategic substitutability in the decision to go to war or attend a protest). This approach is often referred to as \textit{structural methods} or, loosely, \textit{structural estimation}. It uses a formalized mathematical theory, motivated by the political phenomenon of interest, as a foundation for the statistical model. For example, it may use a spatial model of legislative voting as a foundation for a statistical model for estimating legislator preferences or a contest model of war fighting between states to estimate the military returns to being a democracy. 



Structural methods seek to identify and estimate parameters that have a clear interpretation relative to a formal model (see \citealp{Keane10}). This differentiates it from alternative methods. However, contrary to conventional wisdom, structural methods do not have to be computationally intensive or nonlinear, though they can be. They do not have to be derived from a game-theoretic model. There is no inherent tension between structural approaches and experiments, whether lab-based, randomized or quasi-natural ones. In fact, there is structural research that utilizes all these types of data, as we describe in Section \ref{step_by_step}. 

However, by starting from a model motivated by the subject under study, the structural approach has at least four major benefits. First, structural parameters have a clear causal and theoretical interpretation, which is borne from the formal theoretical foundation. This implies that assumptions for their identification and interpretation are clear. Second, by specifying a clear theoretical domain, it can be possible to recover parameters that may not be identifiable (or observable) otherwise. An example being ideologies or preferences, enabling researchers to answer questions that otherwise could not find empirical quantification.\footnote{The difficulty of estimating preferences a-theoretically are discussed in detail in \citet{AbramsonForth} on conjoin experiments.} Third, the extrapolation of estimated effects to new settings is possible, and defined by the scope of the underlying theory, allowing a wide variety of counterfactual experiments and analysis.  Finally, the evaluation of the model itself is of interest. The analyst can ask such questions as: how well does the theoretical model fit the data? Does this model do a better job than other models?  What observations does the model explain well, and which does it not explain well?  Is the model effective at predicting in sample?  Is it effective out-of-sample?

These benefits have substantial payoffs for political science as an empirical and theoretical discipline.  Once one understands that all quantitative analysis is based on models, theoretical structure (i.e., theoretically motivated assumptions) and identification become complementary, and it becomes clear how those interested in the experimental and reduced-form traditions can fruitfully interact with empirically minded theorists.  

Second, providing theoretical structure to empirical analysis forces the researcher to focus on mechanisms, rather than effects, which facilitates engagement with the real world.  Policy oriented empiricists are often skeptical of structural work due to its reliance on several theoretical assumptions. There is a widely held belief that results that follow from what appear to be minimalist or simple assumptions are more useful. The unease can come from skepticism regarding the behavioral and equilibrium assumptions that underlie formal models. However, if the goal is to produce research that can inform policy, it is important to investigate the mechanisms behind causal effects. Because of its lack of connection to behavior or equilibrium effects, reduced-form methods are limited in their ability to identify efficient ways to change an outcome, as well as to understand the consequences of policy changes and counterfactuals that scale \citep{al-ubaydli2017what, Heckman08}.  But we are not completely ignorant of how the world works, theory provides insight that seems wasteful to discard. And this would be especially wasteful for questions and environments dearth of data.  Therefore, a researcher with an eye towards policy implications should be especially interested in pursuing a structural approach.

Finally, and maybe most importantly for political science, a structural approach can facilitate cumulative science.  We have observed that reduced-form studies struggle to generate cumulative knowledge, largely due to our lack of understanding about the external validity of specific results. Frequently, testing general theories solely with reduced-form methods leads us down the path of series after series of contradictory findings, with affirmative results found in one place, negative result in another, in a never-ending process, of which some examples are discussed below. This does not mean either reduce-form result is incorrect. Instead, the problem is our poor understanding of external validity. In principle, although not always in practice, theoretically grounded structural models lend themselves to cumulative knowledge building since they are designed to be externally valid. Then, we would be able to have many scholars developing models, comparing different models, and changing or integrating them in specific attempts to understand features of politics that reduced-form practices cannot discover by construction.

Due to these benefits, structural methods have been growing in prominence in political science and political economy, with a wide set of applications across fields. For instance, it has been used extensively in American Politics, with DW-Nominate \citep{poole1984presidential,bonica2013why} being a salient example, in Comparative Politics to study the role of learning about democracy (\citealp{Abramson20}), voter's use of information in different contexts (e.g., \citealp{Kendall15, Cruz20}), or coalition formation in parliamentary democracies (e.g., \citealp{diermeier03}), in International Relations in the study of conflict and civil war (e.g., \citealp{signorino1999strategic,crisman-cox2018audience,Kenkel21}), among many others. 

However, many find it hard to implement and evaluate structural methods. This is because statistical procedures can be bespoke, and the analyst is often interested in evaluating the whole model or simply quantifying theoretical parameters, rather than testing a parameter's significance.  This means that tools that look more like those found in machine learning, such as non-nested models tests, likelihood ratio tests, joint-significance tests, and cross-validation are often more useful than t-tests. Also, standard errors describe the uncertainty surrounding quantities rather than criteria for rejecting a null hypothesis that was never posed. Section \ref{step_by_step} provides examples of the variety of ways scholars have used statistical techniques to estimate model parameters and propose ways in which researchers, editors, and reviewers can evaluate whether the implementation of such methods in a paper are appropriate or not. We provide many examples among existing works in political science and political economy as we go.

\section{Models Everywhere}

We start from the somewhat trivial observations that all quantitative empirical analysis starts with a model. We could also call these theories, theoretical models, or models of a theory.  In any case, the research assumes a set of untested assumptions about how the world generates data and what is observed.  To fix ideas, consider the following common statistical models.

\subsection{Linear Model}\label{section_linear}
In the classical linear regression model, a researcher posits a linear relationship between an outcome, $Y_i \in \mathbb{R}$, and external variables (e.g., covariates) $X_i \in \mathbb{R}^n$, as:
\begin{equation}
Y_i = X_i'\beta_0 + \varepsilon_i,\label{linear_ex}
\end{equation} where $\varepsilon_i$ is a random variable that is unobserved to the researcher, and $\beta \in \mathbb{R}^n$ are \textit{parameters}. Furthermore, it is assumed that $\mathbb{E} X_i X_i'$ is invertible, and that $\mathbb{E}[\varepsilon_i \mid X_i] = 0$ when $(Y_i, X_i)$ are i.i.d. 

The combination of the linear specification together with the assumptions on the distribution of unobservables and population counterparts to $X_i$ are the model. The researcher assumes the model then aims to estimate the value of the parameter, denoted $\beta_0$, that generates the observed data $(Y_i, X_i)$. Hence, the statistical model is defined as (\ref{linear_ex}) and its assumptions, while the researcher wants to learn the parameter of interest $\beta_0$ from the data. 

\subsection{Logit}\label{section_logit}
In the Logit model, the researcher observes a binary outcome, $Y_i \in \{0,1\}$ and $X_i \in \mathbb{R}^n$, while assuming that:
\begin{equation}
Y_i = 1\{X_i'\gamma_0 + \varepsilon_i \geq 0\},\label{example_logit}
\end{equation} where $1\{.\}$ denotes an indicator variable (i.e., equals 1 if the condition in the brackets is satisfied, and 0 otherwise). 

Again, this model is fully defined by assumptions on population counterparts to $(X_i, \varepsilon_i)$. More precisely, $\varepsilon_i$, which are assumed to follow a (standard) Logistic distribution with CDF $\Lambda(\cdot)$, so that $P(Y_i = 1 \mid X_i) = \Lambda(X_i'\gamma)$, with $\mathbb{E} X_i X_i'$ invertible.

The researcher may wish to learn the value of $\gamma_0$ that generates the data. The model is defined for many values of $\gamma$, but the parameter of interest, $\gamma_0$, is the particular value which induces the observed distribution of $(Y_i, X_i)$, among many possible options. 


\subsection{ATE in the Potential Outcomes Framework}
In the potential outcomes framework, researchers consider a situation where there is a treatment, denoted by $D_i\in \{0,1\}$, that is applied to units of interest and each unit has two potential outcomes,
\begin{equation}
    Y_{Di}=
    \begin{cases}
    Y_{1i} \textrm{ if i is treated}\\
    Y_{0i} \textrm{ if i is untreated.}
    \end{cases}
\end{equation}
The causal effect of $D_i$ on $i$ is
\begin{equation}
    \tau_i=Y_{1i}-Y_{0i}.
\end{equation}
The model is defined such that $D_i Y_{1i}-(1-D_i)Y_{0i}$ is observed, that the treatment assignment is independent of the potential treatment effect $(Y_{1i}-Y_{0i}\perp D_i)$ (e.g., due to randomization) and that there are no spillover effects. \citep[p.140]{cunningham2021causal}. The researcher may be interested, for instance, in the Average Treatment Effect (ATE), $\tau_{ATE} = \mathbb{E} \tau_i$.

\subsection{Ideological Voting in Legislatures}\label{model_ideo_spatial}

Consider a prominent class of models in political science: multidimensional ideological voting. A researcher wants to measure politician ideologies after observing a series of roll-call votes $t=1,...,T$. Each politician $i$ has ideology $\gamma_i \in \mathbb{R}^n$ and chooses to vote ``Yes" ($Y_{i,t}=1$) or ``No" ($Y_{i,t}=0$) on each roll-call depending on whether the alternative policy $x_{t}$ gives them a higher utility than the status-quo policy ($q_t$). 

The preference of politician $i$ for policy $x_t$ is modeled as a random utility composed by a deterministic part, $U(x_t, \gamma_i)$ and a random part $\varepsilon_{i,x,t}$. Hence, politician $i$ votes ``Yes" on roll call $t$ if:
\begin{eqnarray}
    U(x_t, \gamma_i) + \varepsilon_{i,x,t} \geq  U(q_t, \gamma_i) + \varepsilon_{i,q,t}.
\end{eqnarray}

\noindent If $U(k_t, \gamma_i)$ is a quadratic function $\|k_t - \gamma_i\|^2$ for $k_t \in \{q_t, x_t\}$ and $\varepsilon_{i,\cdot}$ is $i.i.d.$ Normal, then this is the model in \cite{CJR04} (absent party effects), \cite{Heckman97}, \cite{rivers03}, among others. If $U(k_t, \gamma_i)$ is a Gaussian function and $\varepsilon_{i,\cdot,t}$ is Normally distributed, then it is DW-Nominate \citep{carroll, boche}.

\medskip

All four examples reflect the essence of any empirical exercise. The researcher's objective is to learn the value of an underlying parameter ($\beta_0$, $\gamma_0$, $\{\gamma_i\}_{i=1}^n$, or $\tau_{ATE}$), from the realizations of a random vector ($(Y_i, X_i, D_i)$ here), having in mind a model (i.e., a specification and series of assumptions, characterizing a ``class" of possible relationships between variables in the data). 

In its most general form, an empirical (statistical) model is a collection of probability distributions over observable variables, derived from a theory of how the world generates the data, such that all possible models are indexed by parameter values and the true distribution is within that class. 
Statistical identification only makes sense \textit{after} defining a model. It is the promised logical conclusion of the statistical (theoretical) proposition for which the model is the hypothesis. 

\section{Theoretical Models and Identification}
\label{identification}
The term \emph{identification} is used in many ways in empirical work (\citealp{Lewbel19}). However, there is only one formal definition, which is related to a \textit{minimal requirement} for a well-defined model.  

Formally, a model is a triple $(\Theta, \gamma, \mathbb{P}_X)$ where $\Theta$ is the set of unobserved parameters and $\gamma$ is a mapping $\gamma: \theta \to \mathbb{P}_X$, where $\mathbb{P}_X$ is the set of all joint distributions over observable random variables $X$.  

 \begin{definition}
     A model $(\Theta,  \gamma , \mathbb{P}_X)$ is identified if and only is for every $(\theta, \tilde \theta) \in \Theta^2 $, $\gamma(\theta)=\gamma(\tilde \theta)$ if and only if $\theta=\tilde \theta$ \citep{athey2002identification}.
 \end{definition}

So $\theta$ is \textit{identified relative to a model} if there is a unique value that rationalizes the distribution of data, i.e., there are not two different parameters $\theta$ and $\tilde \theta$ that, within the model, could induce the same distribution of data, $P_X \in \mathbb{P}_X$.\footnote{There are other types of identification, like partial or set identification, which have similar flavor but do not require the relationship to hold with equality.  See \citet{Lewbel19} for a review of other forms of identification used in empirical analysis.} 

In the linear example above, this requires that only one value of $\beta$ (i.e., $\beta_0$) for the model in equation (\ref{linear_ex}) generates the distribution of the observable $(Y_i, X_i)$.

\subsection{Identification in the Linear Model}
Consider the model in Section \ref{section_linear}. The researcher wants to learn the true $\beta_0$. Under the stated assumptions of the model, we can write:
\begin{eqnarray}
\beta_0 = \left(\mathbb{E} X_i X_i'\right)^{-1} \mathbb{E} X_i Y_i.
\label{linear}
\end{eqnarray}

As we can see, the left-hand side of (\ref{linear}) is the parameter of interest, while the right-hand side is a known function of the distribution of observable data (i.e., of the distribution of $(Y_i, X_i)$). Given the model, \textit{and if the researcher knew the distribution of observables} (including $\mathbb{E} X_i Y_i, \mathbb{E} X_i X_i'$), they could infer $\beta_0$ for sure. That is, there is no other value of $\beta$ that could generate the distribution of the data. Hence, $\beta_0$ is identified. 

Notice that if $\mathbb{E} X_i X_i'$ failed to be invertible (i.e., $X_i$ suffered from multicollinearity), then $\beta_0$ would not be identified. Indeed, there would be multiple values of $\beta$ that could generate the same data.\footnote{For instance, if $Y_i = \alpha + \beta_0 X_i + \varepsilon_i$, with $Var(X_i) = 0$, then this model would lead to the same joint distribution of $(Y, X)$ as $Y_i = \tilde \alpha + \varepsilon_i$, with $\tilde \alpha = \alpha + \beta_0 \mathbb{E} X_i$.} Note that the multicollinearity is a problem with the data generating population model, not with the sample.  Samples play no role in identification, whether a parameter is identified or not is a statement about the nature of the theory of the data generating process and what is observed in hypothetical infinite samples. 

\subsection{Logit}\label{section_logit_id}
We now revisit Section \ref{section_logit}. Under the stated assumptions,
\begin{eqnarray}
\label{logit}
\mathbb{E}[Y_i\mid X_i] &=& \Lambda(X_i'\gamma_0) \notag \\
\Lambda^{-1}(\mathbb{E}[Y_i \mid X_i]) &=& X_i'\gamma_0 \notag \\
\mathbb{E} X_i X_i' \gamma_0 &=& \mathbb{E} X_i \Lambda^{-1}\mathbb{E}[Y_i \mid X_i] \notag \\
\gamma_0 &=& (\mathbb{E}X_i X_i')^{-1} \mathbb{E} X_i \Lambda^{-1}(\mathbb{E}[Y_i \mid X_i]).
\end{eqnarray}
Again, the right-hand side is a known function of the moments of $(Y_i, X_i)$, so $\gamma_0$ is again identified.

By comparison, suppose that $\varepsilon_i$ was Logistic, but with scale parameter $\sigma \neq 1$. Then, the model $Y_i = 1\{X_i'\gamma_0/\sigma>\varepsilon_i/\sigma\}$ would generate the same distribution of $(Y_i, X_i)$ as the one above. Hence, the researcher observing the distribution of $(Y_i, X_i)$ cannot know if the data came from the first model (with parameter $\gamma_0$), or from the second model (with parameter $\gamma_0/\sigma$). Hence, $\gamma_0$ would not be identified if the distribution of $\varepsilon_i$ was unknown.

\subsection{Potential Outcomes and ATE}  In the potential outcomes framework, we can write the observed $Y_i$
\begin{equation}
    Y_i=Y_{i0}+(Y_{i1}-Y_{i0})D_i + \sum_{j\neq i} \rho_{ji} D_j,
\end{equation}
where the first two terms define the observed outcome from $i$ and the third is the spillover effect of the treatments on the other $j$ observations.  Assuming no spillovers, thus that $\rho_{ji}=0$ for $j$ and $i$, we have that 
\begin{gather*}
    \mathbb{E}[Y_i|D_i=1]-\mathbb{E}[Y_i|D_i=0]=\mathbb{E}[Y_{i1}|D_i=1]-\mathbb{E}[Y_{i0}|D_i=1] +\mathbb{E}[Y_{i0}|D_i=1]-E[Y_{i0}|D_i=0].
\end{gather*} 
The latter term is the selection effect, which is zero when $Y_{1i}-Y_{0i}\perp D_i$.  Homogeneity and the linearity of the expectation operator give us that
\[\mathbb{E}[Y_i|D_i=1]-\mathbb{E}[Y_i|D_i=0]=\mathbb{E}[Y_{i1}-Y_{i0}|D_i=1]= \mathbb{E}[\tau_i].\]

If the no spillovers assumption were violated, then there would be many $(\rho, \tau)$ pairs that would produce the same difference in observed $Y_i$ and the model would be unidentified.

\subsection{Identification of Ideologies}

We refer the reader to \cite{rivers03} for a careful discussion and proof of identification for a general dimension of preferences $n$ and quadratic preferences $U(\cdot)$ covered in Section \ref{model_ideo_spatial}. While a similar intuition to the previous sections can be applied, the proofs are more subtle given the non-linearities and the large number of parameters (e.g., $n$ for each politician, plus $x_t, q_t$ for every roll-call). 

Meanwhile, \cite{CKT22} presents a careful discussion of identification in 1-dimensional DW-Nominate, and the challenges with identification in 2-dimensional DW-Nominate. (See Appendix B in that paper).

\subsection{Important Messages on Identification}

In this discussion of the notion of statistical identification, it is worth drawing attention to some key points.

First, identification is always relative to a model. That is, there cannot exist an argument of identification that holds ``atheoretically" or simply by ``data". This is because the model allows us to define the parameters we are interested in, and requires us to be explicit about its underlying assumptions. 

Indeed, if one changed the theoretical modeling assumptions, there would be no guarantee that the parameter of interest would be identified. For instance, assume that we changed equation (\ref{linear_ex}) in Section \ref{section_linear} to include a new random variable on the right-hand side. Then, the original $\beta_0$ may fail to be identified. Similarly, $\beta_0$ would fail to be identified if we changed the exogeneity assumption to $\mathbb{E}[\varepsilon_i \mid X_i] \neq 0$. There is no reason why identification of a parameter would hold in a different model.

Second, identification is always about population counterparts, and \textit{never} about samples. Suppose that a researcher knew everything they could ever wish for from the variables she observes (e.g., the whole distribution of $(Y_i, X_i)$ in Sections \ref{linear}-\ref{logit}). In the notation above, they know $P_X$ and, hence, all moments of the observable data. Identification asks us whether the researcher could then infer the true value of the parameter from this known distribution. In other words, could they uniquely pin down which parameters generated the data we observe? If not, then the model is ``ill-defined": \textit{not even in the ideal world could you know which parameters generated that data.}

Of course, we do not live in such an ideal world. We do not observe $P_X$, we observe a sample of $(Y_i, X_i)$.  However, identification comes before even thinking about such a sample: if we could not find out the true parameter \textit{even if we knew the true population}, how can we expect to estimate anything close to $\theta_0$ having only a sample instead?

Finally, in any empirical work, the researcher's goal is to identify their parameter of interest. In many models, some parameters might be identified while others are not. In this case, the researcher must just be sure that their parameters of interest are identified. It is not strictly necessary \textit{for all} parameters to be identified, if some of them are useless to answer the research question.\footnote{One example is that Probit models are not identified when the error term has variance different than 1. However, the Average Partial Effect, $APE = \int \frac{\partial \mathbb{E}[Y_i \mid X_i = x_i]}{\partial x_i} dF_X$, where $F_X$ denotes the marginal distribution of $X_i$, is identified. See \cite{Marshak53}, \cite{Heckman10} for a discussion.}

\subsection{Estimation}
Estimation is the secondary process of quantifying theoretical parameters from samples. An estimator $\hat \theta$ for $\theta_0$ is a function of the observed sample. In the linear example above, an estimator is any function of the sample $(Y_i, X_i)_{i=1}^n$. That is it. A common estimator in the linear model is the Ordinary Least Squares Estimator (OLS). But that is not the only one. For instance, a Maximum Likelihood Estimator (MLE) could be used, and so could an estimator that is just the first observation of $X_i$. Different estimators will perform ``better" or ``worse" depending on its assumptions. OLS is often used because of its excellent properties for the linear model, but such properties do not hold in other models.

\begin{figure}
\caption{Identification and Estimation}
    \centering
    \includegraphics[width=0.6\textwidth]{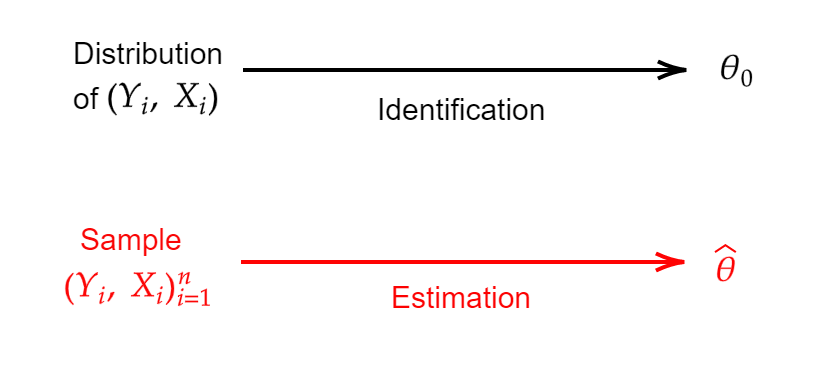}
    \label{fig:model_id_est}
		 \parbox[c]{6.2in}{%
{\footnotesize{}Notes: A model specifies how parameters generate the distribution of $(Y_i, X_i)$. Identification asks how can we learn $\theta_0$ from the distribution of $(Y_i, X_i)$. Identification is about population features. When we have a sample (i.e., realizations of the distribution of $(Y_i, X_i)$), then we try to estimate $\theta_0$.}\bigskip{}
 }%
\end{figure}

We remark on two important aspects in this definition. First, an estimator is always defined relative to a parameter. Otherwise, what is the estimator seeking to quantify? Second, an estimator is based on the sample, not the population. It is a function of realized data, not the ``ideal world" of identification. Hence, estimation can only come after identification. After all, it only makes sense to estimate $\theta_0$ once we know that it is statistically identified. Otherwise, (i.e., if we could not pin $\theta_0$ down even if we knew everything we wished about the observable data), the estimator has no hope of providing a good approximation of $\theta_0$. Figure \ref{fig:model_id_est} below provides an illustration of these concepts.

It is common for authors to confound identification and estimation in their presentation.\footnote{See, for example, \citet[p.581]{burke2015climate}. One common such mistake is to say ``estimate an OLS model". This confounds the estimator (OLS), with the model (linear) and its target parameter ($\beta_0$ in Section \ref{section_linear}) that is being estimated.}  While usually harmless in terms of the interpretation of empirical results, this can lead to confusion about what assumptions imply the model and observable population moments make the calculation of a parameter possible versus what is necessary for a sample to provide a reliable estimate of this quantity.  

\FloatBarrier

\section{Theory, Interpretation, and Extrapolation\\ \normalsize (or, the Limits of Inference Without Theory)}\label{section_theory}

The previous sections show how the technical elements of empirical work requires models.\footnote{The title of this section is drawn from the excellent \cite{Wolpin13} book, from which we draw inspiration.} This conflicts with the widespread view that ``letting the data speak for itself" or that a ``data-driven" analysis is possible. It can be very appealing to think that one's work does not require assumptions, and that one is free from theory in their interpretation of the results. Unfortunately, this is not the case.  In addition to providing a technical foundation for a quantitative empirical exercise, theoretical models, whatever they may be, impart meaning and scope to the analysis.

\subsection{Causality is a Theoretical Claim}\label{section_causal}
An early discussion of the deep role of theoretical models goes back to a point made at least since \cite{Koopmans47}, who argued one cannot even measure variables without theory. This point also extends to causal claims.

The ``credibility revolution" made an enormous impact in empirical work in the social sciences. In particular, it brought much needed attention to statistical identification and the need for care in empirical work. However, a byproduct of its popularity has been an interest in finding ``causal" effects with little attention the meaning of the estimand \citep{lundberg2021whata}. Causality, derived from an equilibrium model or the Rubin causal model, only exists relative to the theoretical model. 

For example, return to the Logit model and equation (\ref{logit}),
\begin{equation}\gamma_0 =(\mathbb{E}X_iX_i')^{-1} \mathbb{E}X_i \Lambda^{-1}(\mathbb{E}[Y_i|X_i])
\end{equation}
where $\Lambda(\cdot)$ is the Cumulative Distribution Function and $X_i$ and $Y_i$ are random variables.  

This is a formula that maps numbers to numbers and has no intrinsic meaning.  As show above, it can be derived from a latent variable model with a logistic distribution function. In that context, $Y_i$ is an indicator that equals 1 when $X'_i \gamma_0 \geq \varepsilon_i$ and $0$ otherwise. $\gamma_0$ equals the coefficient that describes a unit change in $x$ causes a change in the log-odds of the outcome. 

But one could also come to this formula in at least two different ways. Suppose a researcher were interested in individual choice, and a decision problem where an agent had to choose between two alternatives.  A natural first step would be to consider a theory of decision-making where (i) the agent selects alternatives with a probability that is monotone in the alternative's value and (ii) satisfied the principle that the consideration of new alternatives does not change the relative likelihood of choosing between two given alternatives, the latter being a form of statistical independence of irrelevant alternatives. \cite{luce59} shows that there is an axiomatic theoretical connection between this theory of decision-making and the Logit model.  Specifically, if the value of an alternative is given by $v(x)=u(x)+\varepsilon$, and $\varepsilon$ has an Extreme Value Type 1 distribution, such a theoretically derived choice rule is consistent with probabilistic utility maximization and is equivalent to Logit. Here the $x$'s are attribute features and $\gamma_0$ is their effect on the valuation of the alternative. That is, the parameter is the effect of $x$ on the decision-maker's utility for selecting an alternative.

In the context of strategic contests over a disputed good, the analysis in \citet{Kenkel21} shows that the standard Tullock model, with a natural parametrization of the marginal returns and marginal costs of effort, generate equilibrium strategies which are equivalent to a Logit probability model of winning the contest. That is, their strategic theoretical framework implies this same equation. In this case, $\gamma_0$ are marginal returns to effort from the observed factor.  

The point of this example - just like the first example in the Introduction - is that any time we want to travel the path between population parameters and meaning, the theoretical model is serving a second important purpose, a point carefully elaborated in \cite{Wolpin13}.  It could be a statistical model, the Rubin causal model, or the equilibrium of a game.   In any case, the meaning of $\gamma_0$ follows from the application of an underlying theoretical framework.

An accurate way to understand structural methods' then is the aim to identify and estimate model-specific parameters in the light of a substantive theoretical model. This may include preferences (e.g., ideologies), measures of welfare, parameters governing agent behavior (e.g., the magnitude of strategic substitutability in the decision to go to war), or policy evaluation which depends on a clearly defined model (e.g., the implementation of a policy that has never happened before).

This is not just a philosophical point. One salient example within political science has been the discussion about whether extreme events (e.g., shark attacks, or college football results) affect voter behavior \citep{achen2017democracy}. A positive finding is sometimes interpreted as evidence of voter irrationality - that is, why should voters change their opinions about politicians when faced with independent events?\footnote{See \cite{Graham22} and \cite{Fowler22} for recent results and an overview of this debate.} However, we can use theory to take a step back and guide our interpretation of this debate, an approach pursued in \cite{Fowler22}. In particular, it is completely possible that the presence of extreme events affecting voter behavior may still be consistent with Bayesian learning (e.g., through an extreme event informing voters about politicians' behavior in relevant events) - see \cite{Ashworth18}. Hence, the observation of linear model's estimates \textit{absent theory} is not sufficient to inform us about what is being recovered. Rather, the researcher must stipulate whether it is reasonable to assume that even a rational voter could learn about an incumbent's type from shark attacks.

\subsection{External Validity and Extrapolation are Theoretical}

Current research in statistics is also concerned with the problem of external validity and extrapolation confronting much statistical analysis \citep{findley2021external, egami2022elements,hartman2021generalizing}.  Explicitly using substantive theory can solve both of these problems.  When it comes to external validity and extrapolation, what researchers want to know is to what population the results naturally generalize?  

Theoretical models are constructed explicitly to represent a class of well-defined events.  For example, a structural model of elections in advanced democracies applies to advanced democracies, a structural model of interstate wars applies to interstate wars, and a structural model of congressional voting applies to congressional voting.  Estimates from structural models are also deep parameters of the underlying theory.  This allows the research to take those parameters values and utilize them in related but new environments.  Extrapolation makes sense once the quantities estimate are critical primitives of the theoretical framework.  

\section{Structural Methods: A Primer}\label{section_structural}

The previous sections lay what is needed to execute and interpret any empirical analysis. None of these arguments or claims are limited to research that is structural. We now overview structural methods themselves. To do so, it is convenient to revisit its formal (historical) definition of structural analysis, relating it to the current use of the terms ``structural" and ``reduced-form". 

\subsection{``Structural" and ``Reduced-Form" Methods: A False Dichotomy}

Structural methods are often contrasted to reduced-form methods. The latter are now associated with research designs such as Difference-in-Differences (DiD) and Regression Discontinuity Designs (RDD), etc, while the former are more prevalent in papers with formal models.\footnote{However, such definitions are imprecise. Research designs like Randomized Control Trials, RDD's, DiD etc. also need to define a theoretical model (why are we changing $X$?), which imposes further assumptions (e.g., randomized assignment of $X_i$, or parallel trends). Each model can then identify a parameter of interest (e.g., the Average Effect of a Policy), which is estimated using a specific estimator (which may be OLS, nonparametric, etc.).} This has led to a division of camps, where researchers tend to specialize in one or the other, with little crossover. It is also sometimes associated with antagonism (see \cite{Angrist10}), as they are thought to be opposites.

However, such a distinction is wrong, both in terms of their definitions, and in their historical use. So that we can survey structural methods, it is important to then define what its sibling, ``reduced-form" is. Its definition is clearest in the following simple example of a Richardson arming model from international relations \citep{richardson2012arms,dunne2007chapter}.

Assume there are two agents $i=1,2$, who choose how many arms to secure $Y_i \in \mathbb{R}_+$. Further, assume that each agent has a characteristic $X_i \in \mathbb{R}$ (unidimensional, for simplicity). A version of the Richardson model posit the following model of the joint decision for arming:
\begin{eqnarray}
Y_1 = \alpha_1 + \beta_1 Y_2 + \gamma_1 X_1 + \varepsilon_1 \notag \\
Y_2 = \alpha_2 + \beta_2 Y_1 + \gamma_2 X_2 + \varepsilon_2 \label{IR_struct},
\end{eqnarray}
so that $\Theta = (\alpha_1, \alpha_2, \beta_1, \beta_2, \gamma_1, \gamma_2)'$ are parameters, and $\varepsilon = (\varepsilon_1, \varepsilon_2)'$ are mean-zero unobservables that are independent of $X = (X_1, X_2)'$. The empirical researcher observes $(Y_i, X_i)_{i=1,2}$.

Equations (\ref{IR_struct}) and their associated assumptions are called \textit{the structural model}. This is because they outline the original model posed by the researcher, clearly delineating how such relationships depend on the fundamental, or \textit{structural} parameters, $\Theta$.

However, one can rewrite the same model (\ref{IR_struct}) by inverting the system so that all dependent variables $Y = (Y_1, Y_2)'$ are on the left-hand side of the expressions. This yields:
\begin{eqnarray}
Y_1 = \frac{\alpha_1+\beta_1 \alpha_2}{1-\beta_1\beta_2} + \frac{\gamma_1}{1-\beta_1 \beta_2} X_1 + \frac{\beta_1 \gamma_2}{1-\beta_1 \beta_2}X_2 + \frac{\varepsilon_1+\beta_1 \varepsilon_2}{1-\beta_1\beta_2} \notag \\
Y_2 =\frac{\alpha_2+\beta_2 \alpha_1}{1-\beta_1\beta_2} + \frac{\beta_2 \gamma_1}{1-\beta_1 \beta_2}X_1 + \frac{\gamma_2}{1-\beta_1 \beta_2} X_2 + \frac{\varepsilon_2  + \beta_2 \varepsilon_1}{1-\beta_1\beta_2},
    \label{IR_reduced}
\end{eqnarray}
preserving the same assumptions on $\varepsilon$ and $X$. Equations (\ref{IR_reduced}) are called the \textit{reduced-form} equations, because they reduce the structural model. Indeed, ``output" variables are now placed in the left-hand side, while ``input" variables are all on the right-hand side. Hence, the coefficient $\frac{\gamma_1}{1-\beta_1 \beta_2}$ on $X_1$ in the first equation is the reduced-form effect of $X_1$ on $Y_1$.\footnote{In the terminology of \cite{Heckman10}, we have the ``internal" variables $Y$ written as a function of the ``external" variables, $(X, \varepsilon)$. When $(X, \varepsilon)$ are independent, then we can say the left-hand side has the dependent variables, and the right-hand side has the independent variables.} This is the only formal definition of ``reduced-form" we are aware of, and goes back at least to the Cowles Commission work in the 1940's. 

Note that a researcher interested in the effect of $X_1$ on $Y_1$ does not need to identify all of $\Theta$: she only needs to identify the \textit{reduced-form parameter} $\frac{\gamma_1}{1-\beta_1\beta_2}$, and not $\gamma_1$ separately from $\beta_1$ and $\beta_2$ as posited in (\ref{IR_struct}). At the same time, its causal interpretation does not come from the reduced-form expressions, but rather from the original model itself.

Hence, depending on the question of interest, the researcher can rely only on the reduced-form of the model, rather than on its structural counterpart. This is what gave rise to the current use of the terminology. However, for other questions, the researcher might be interested in the specific values of $(\beta_1, \beta_2)$, for instance, which will require them to prove they are separately identified.

The fundamental difference between (\ref{IR_struct}) and (\ref{IR_reduced}) is that the former is the theoretical data generating process and the latter is an accounting that respects the theoretical structure, but arrays endogenous variables on the left and exogenous variables on the right. Writing the reduced-form as (\ref{IR_reduced}) does not depend on whether a researcher is structural or reduced-form. Rather, the choice between (\ref{IR_struct}) and (\ref{IR_reduced}) is whether to display the model's dependence between external and internal variables, and possible challenges in identifying the parameters $\Theta$. Because agents 1 and 2 interact, their characteristics impact their choices, which influences the choices of the opponent.

As we can see, the comparison between the traditional definitions of structural and reduced-form are deeply connected. The more precise distinction is that a structural work wants to identify and estimate fundamental parameters of this model - i.e., the (whole or a subset of) vector $\Theta$. In this case, authors must fully specify the model, which includes (\ref{IR_struct}) as well as the further assumptions (e.g., random vector $\varepsilon$ being mean zero). Then, the author proves the identification of these parameters under this model, and can proceed to discuss estimators. Meanwhile, an author interested in the total effect of $X_1$ on $Y_1$ only needs to identify and estimate a parameter, which is the composition of $\gamma_1, \beta_1, \beta_2$. But in doing so, they are still assuming the original statistical (and formal) model  they are simply extracting different information from the data.

\subsection{The Modern Use of the Term ``Structural"}

With this example in mind, the term structural estimation is used to describe researchers want to identify and estimate model-specific parameters (i.e., parameters defined relative a clearly laid out theoretical model with its associated assumptions). This is, of course, motivated by their research question.\footnote{Some researchers consider the terminology ``structural" estimation to be a misnomer since, as we saw above, any estimation only makes sense after some (structural) model has been put in place. However, we keep the definition that is most appropriate for its use in the literature.}

As we can see, there is nothing in the definition of structural methods that requires non-linear models or computationally intensive approaches, there are structurally estimated models that are linear and estimable by simple methods (e.g., Ordinary Least Squares - OLS) \citep{Heckman97, smith1989models}. Furthermore,  experimental or quasi-experimental designs either can aid in identifying parameters within structural models. For example, a Randomized Control Trial (RCT) can be used to identify voter beliefs and preferences, once it is carefully used within a formal model (e.g., \citealp{Kendall15, Cruz20}). See \cite{toddwolpin20} for an extensive discussion. Discontinuities and quasi-experimental methods can be used in structural methods too. For instance, \cite{Martin17} explicitly use the random ordering of the listing of Fox News Channels to help identify preferences for like-minded news. 

Hence, as \cite{Keane10} writes, \textit{``the distinction between reduced-form methods and structural ones are not in the number of assumptions (as the model above shows), but rather in how explicit they are about such assumptions, and their objectives".}

Often, structural papers seek to identify different parameters (whether ideologies, preferences, productivities, etc.) than non-structural ones (e.g., average effect of a policy), and rely on a formal description of the theoretical environment the researcher uses to interpret their estimates. Hence, relative to the bulk of empirical work in political science, structural models require researchers to stipulate the formal theoretical model they have in mind and outline the assumptions, clearly define the target parameter, describe how they map to data, and explain how their parameters retain their interpretation in counterfactual and new environments (as long as the original theoretical model holds). 

In doing so, researchers can use their estimates to answer important questions which would not be possible otherwise, as we discuss in the next section.

\subsection{The Types of Questions of Interest}

Should a researcher use structural methods? This fundamentally depends on the question of interest. Here, we outline our views on what types of research questions structural methods are best suited. 
In our view, there are three such classes of questions. 

The first class is research questions that look to quantify theoretical parameters (i.e., those which are only defined relative to a theoretical model). For instance, DW-Nominate and other measures as estimates of the ideology of members of Congress (\citealp{poole1984presidential, bonica2013why, Heckman97, CKT20, CKT22}, among many others) or audience costs \citep{crisman-cox2018audience,kurizaki2015detecting}. Ideologies are preferences, which can only make sense relative to a model which defines preferences and individual's actions (in DW-Nominate, votes on bills) and audience costs are important when they are unobserved. This class of questions also includes measures of strategic complementarity or substitutability of political protests (e.g., \citealp{Cantoni19}), the weight that politicians give to reelection concerns (\citealp{ilm22}), the extent of spillovers in the provision of public goods along a network (\cite{Acemoglu15}), voter preferences from aggregate data \citep{Iaryczower22,rekkas07,ujhelyi21}, the role of preferences for like-minded news in driving media slant (\cite{gentzkow10}), welfare effects (e.g., consumer surplus), which are model dependent, and others.  In each case, the goal is to estimate a theoretically important parameter in a way that can only be done using structural techniques.\footnote{This can further include using theory to help measure behavior that is hard to observe, such as connections among legislators in the U.S. Congress, as \cite{battaglini21, cjt}), or how competition in lobbying affects policy approval (e.g., \cite{kang2016policy}) and government contracts (\cite{cox22}).} 

The second class of questions are counterfactual scenarios or hypothetical policy interventions that explicitly require extrapolation of results to new environments. This can include questions which extrapolate policies beyond observable data (e.g., whether a campaign contribution limit in Colombia or Brazil - see \citealp{Gulzar21, Avis22} - would work similarly elsewhere). Extrapolation requires some information that tells the researcher that their results should hold elsewhere. This role is played by a theoretical model that the researcher deems reasonable to be applied in both settings.

Similarly, structural methods are particularly well-suited for evaluating policies that have never been implemented before or that are unobserved. For instance, one could not evaluate the effects of a minimum wage increase on minority voting behavior, when the increase in minimum wages has not been observed in the data before. Absent a formal model, it is unclear how to estimate the effect of such a policy, as it is beyond the data's support. Rather, such policies can only be evaluated by using tightly connected theory: whether to extrapolate other observed increases (e.g., due to parametrizations), or re-computing equilibria in strategic settings (where simple extrapolations of existing data may fail to be an equilibrium in the counterfactual setting).\footnote{Such views are well summarized in \cite{Heckman08}'s categorization of policy evaluation questions, which permeate empirical work. He splits such questions into three broad categories. First are questions that evaluate historical interventions on observable outcomes. For such questions, existing reduced-form methods may suffice, as there may be no need to define formal models beyond the statistical model itself (i.e., specification and model assumptions needed to recover treatment effects). Second are questions about interventions implemented in one environment, and whether they can extrapolate to another. Clearly here, there needs to be more theoretical assumptions than in the first questions. In particular, one has to define what parts of the theoretical model in one environment remain invariant in the next. Finally are questions about forecasting interventions never observed in the data. This typically requires a much more thorough and detailed theoretical model.} 

Such an exercise can be found in \citet{iaryczower2012value}.  In their analysis of the U.S. Supreme Court, they use counterfactual simulations to compare the performance of the court as it exists to a counterfactual court where ruling against the Defendant requires the unanimous consent of the justices.  The consequence of this rule change is not obvious because both the nature of the information justices have, and their political biases, influence their votes and the quality of the court's collective decision-making. We know that in a collective decision-making environment with a heterogeneous group of individuals in terms of preferences and abilities, majority rule does not always outperform unanimity rule. The results for the U.S. Supreme Court, however, show that unanimity rule leads to a larger probability of error than simple majority rule.  Such an analysis is only possible given the structural model.

Fourth, structural techniques can be very useful for distinguishing between competing theories or models of a single class of events.  In many areas of political science, we have several models of the same phenomenon. Structural estimation can be viewed as a form of statistical model calibration, forcing the model to interact with the data allows the research to ask questions about model fit and compare the performance of different models.  A good example of this kind of work is \citet{Francois15}.  In that article, the quantification of the model of power-sharing allowed for a direct comparison with several alternative theories, like the ``Big Man" theory of power centralization, and make quantitative claims about how well the various theories fit the data. Another example is in \cite{kang22}, where the authors can estimate and compare the roles of (i) information, (ii) seller characteristics, (iii) administrative hurdles, and (iv) corruption in explaining why there is so little competition for government contracts.

\section{Building a Structural Model from the Ground-Up}\label{step_by_step}

Admittedly, the previous sections related to ``structural methods" in a very abstract way. We rectify this by overviewing the steps which, in our view, could be followed to build and estimate a model structurally. In doing so, we kindly remind the reader that the decision to use a structural method depends first and foremost on their research question.

\subsection{First Step: What is the Research Question and Parameter of Interest?}
Whether a fully specified structural model is necessary depends on the object of interest. Hence, the first step for any structural model is to carefully define the target parameters. 

For example, one might ask what determines the effectiveness of coalitions in war?  Such a question implies some relationship between characteristics of actors and war outcomes. In particular, what characteristics influence the effectiveness of a fighting force?  What kinds of attributes make war fighting mores costly?  How much effort can a country expect from their partners and themselves?

Note that the definition of the target parameter is more specific than ``the effect of policy Z" or a ``causal effect". There are infinite parameters that could fit such descriptions. Rather, we encourage the researcher to be specific: upon which population are we studying the effect of policy Z on? Do we care about average effects, effects on the poorest, on the richest? Or effects of a policy on rural versus urban communities? In addition, which causal effect do we want? Partial or general equilibrium or holding some actions fixed? Accounting for externalities or not? Etc.
Only after this is clearly defined relative to the political scientist’s research question, can one make progress.

Again, we want to emphasize that this step is the same regardless of the empirical or theoretical approach. 

\subsection{Second Step: What (Formal) Model Best Approximates the Empirical Set-up?}

Given a clear objective, it is now up to the researcher to define the formal environment of their study, where the parameter in Step 1 has meaning. This includes who are the actors in this environment, whether it is strategic or not, the information structure (what agents know and what do they not know), the timing (e.g., is the environment is static or dynamic) and so forth.

The end-goal of this step is for the formal model to provide predictions of observable data as a function of underlying model-parameters. For instance, in the study of political polarization in Congress, formal models predict a politician's decision to vote Yea or Nay on a bill as a function of their ideologies (e.g., DW-Nominate, as discussed in Section \ref{model_ideo_spatial}), while others further allow this to depend on party discipline parameters (e.g., \citealp{CJR04,CKT20,CKT22}). Some models have quadratic preferences (e.g., \citealp{Heckman97}), while others are Gaussian (DW-Nominate). Nevertheless, all of them generate observable predictions as a function of model-specific parameters, as we showed in Section \ref{model_ideo_spatial}.

The decisions on the model environment are best guided by theory, data, institutional knowledge and internal consistency, among others. For instance, is it reasonable to approximate the role of deliberation among Supreme Court justices in their decisions as an environment where such Justices have complete information of each other’s preferences (given they work together for decades)? Or, is it better reflected by them having private information (e.g., \citealp{Iaryczower18})? Is there a pattern in the data that suggests private information is a more appropriate assumption – e.g., a different correlation structure of outcomes or, alternatively, does the theoretical model with private information induce equilibrium outcomes that better resemble patterns in the raw data?

Sometimes, that decision is simpler as the parameters can be derived from more widely used theoretical models. For instance, \cite{Acemoglu15} study public goods provision across municipalities in Colombia. They are interested in the extent of such spillovers: municipalities can choose costly public good provision, but they can also free ride on others'. This decision is modeled as a strategic choice on a network, whose equilibrium map informs estimation using the outcome (data on public provisions), a network (geographical proximity of municipalities) and further municipal-level data.  

In the example of coalition war fighting referenced in Section \ref{section_causal}, \citet{Kenkel21} join together the canonical bargain model of crises and a contest function between teams.  Equilibrium analysis shows that the probability of a coalition winning a war is a generalized Logit when there are more than one member on each side, but a simple Logit for wars between two countries.

Given structural work takes the formal model to the data, such decisions about the environments are particularly important. Even if an environment is sufficient to identify the parameter of interest (Step 3 below), it is possible that the fit or empirical performance is worsened by not taking such characteristics into account (see Step 4 below).

\subsection{Third Step: What Conditions are Required on the Model so that the Parameters are Identified (Given Data)?}

By this step, the researchers should feel comfortable with the objective and the general approximation of their model to the environment. They should have also obtained the model's predictions as a function of parameters: e.g., equilibrium relationships that relate observable data and only depend on parameters. The first two steps resemble those in any other empirical work. It is the third step where the structural approach differentiates itself from formal theory.

It is no longer enough for the formal model to obtain a theoretical prediction which maps into a target parameter. We now have to prove whether the parameters of interest are identified. Because each formal model may be different than one another, there is no clear recipe to implement. However, we overview some existing paths, keeping in mind the definitions in Section \ref{identification}.

Typically, the researcher uses the model's predictions of observable outcomes, written as functions of other observable data. For instance, in models with strategic interactions, the equilibrium outcomes (e.g., going to war or not) will be functions of parameters, as well as characteristics of the players of the game. The outcomes and characteristics are known, and we have to find conditions on unobservables and parameters for identification. Examples of complete proofs include \cite{CKT20,CKT22,Kawai21}, among others.

In doing so, the researcher often finds that additional restrictions need to be imposed: e.g., a moment condition, a normalization (the mean of a distribution shock has to be zero), an independence restriction (e.g., individual types have to be independent from other observables), or a parametrization (the distribution of shocks will be Logistic, or preferences will be a linear function of observable characteristics).\footnote{We note that a normalization could then change the interpretation of the identified parameter: e.g., a normalization may seem innocuous at first, but may change the interpretation of a parameter of a counterfactual effect – see \cite{Aguirregabiria14}.} Other times, the choice may be motivated by tractability and computational considerations, which is why Logistic is a popular choice. Again, the choice of whether that is reasonable or not depends on whether (i) seems reasonable in that context (i.e., are agent types likely independent from the policy?), (ii) it changes the interpretation of a parameter, (iii) would miss a key feature of the data.

It is often even easier to find out whether a parameter fails to be identified. This is simply whether there is not enough information in the data to differentiate the parameters' roles in the model. For instance, if two parameters always appear multiplying each other in the model, then they play observationally identical roles despite their differences. They cannot be separately identified, and a normalization will be necessary.\footnote{Even when identification fails, the model can still be informative. For instance, multiple equilibria imply the multiplicity of solutions, given the same observable data. Hence, parameters may fail to be pinned down in such settings. Nevertheless, one may still find that outcomes are positive regardless of the equilibrium. The literature considers such cases in partial identification – which can be due to incomplete data or model, but may still be useful. We do not review such approaches here, deferring the reader to works as \cite{Molinari20}.}

Hence, the researcher needs to pick reasonable assumptions to their problem so that identification can be obtained. How to measure whether an assumption is ``reasonable" depends on theory and on the question. Some identification restrictions are fundamentally not testable: this includes structure on the distribution of unobservables. This is similar to the typical ``exclusion" restrictions in Instrumental Variable models. Nevertheless, they may be reasonable given existing theoretical work, or qualitative or quantitative evidence. Other assumptions are easily testable. For instance, whether a specified correlation exists in the data.

In our running example of the effectiveness of coalitions in war, covered in Section \ref{section_causal}, we know the model is identified because the Logit model it identified, as proven in Section \ref{section_logit_id}.  

\subsection{Fourth Step: Taking it to the Data.}
Now the researcher has formally proven statistical identification of their parameter of interest. They consider that their formal model captures  key parts of their problem, and that their further assumptions are reasonable. Now, they are ready to take their model to the data.

Hence, it is time to choose an appropriate estimator – i.e., an estimator with reasonable properties for their problem (e.g., consistency and asymptotic normality). Such a choice inherently depends on their resulting model. If it is linear, and additively separable (e.g., the linear model in Section \ref{section_linear}), then Ordinary Least Squares is a relevant candidate. This is because the latter has numerous advantages: it is consistent, has a closed form solution, it is asymptotically normal and, under homoskedasticity, attains lowest variance among linear unbiased estimators.

However, most structural models do not end up being linear and additively separable. For nonlinear models, the choice depends on the structure of nonlinearity. For example, if the equilibrium conditions that were used for identification can be written as moment conditions, then a candidate estimator is the Generalized Method of Moments Estimator (GMM). Under regularity conditions, such an estimator will be consistent and asymptotically normal (\citealp{Hansen82}). When there are more moment conditions than parameters, then one can provide weights to make it efficient within a reasonable class of estimators. The GMM estimator is semi-parametric, as it only uses moments of data (expectations), but not the whole distribution of the data. This is convenient if the researcher has not imposed parametric distributions on the unobservables of their model. 

However, in other models, researchers may have already imposed further parametric assumptions: either for identification, or because they help capture important patterns in the data (e.g., thick tails), because they are useful to deal with selection or missing data (e.g., \cite{Heckman79}) or because they are important for theoretical purposes (\citealp{CKT20} use normally distributed shocks, which guarantee that a monotone hazard rate condition for equilibrium uniqueness is satisfied). In such cases, Maximum Likelihood Estimators become an ideal candidate. This is the case for the Logit example in Section \ref{example_logit}, which also applies to our main example in Section \ref{section_causal}, as a Logistic assumption has already been imposed. And, MLE achieves the Cramer-Round bound for lowest variance within a reasonable class of estimators (asymptotically unbiased).

Maximum Likelihood Estimation, though, may be very computationally intensive. This is the case when one must integrate over multidimensional unobservables, and such integrals may not have a closed-form. In such cases, simulation methods or indirect inference may be convenient: a researcher tries many different parameter combinations, simulates the model for each one, and checks which ones fit the data best. The properties of each estimator, and many others, are well reviewed in statistics textbooks, such as \cite{Wooldridge10}. With such estimators in hand, researchers can evaluate their underlying hypothesis through appropriate inference - e.g., hypothesis tests using the asymptotic distribution of those estimators.

\subsection{Fifth Step: Evaluation of the Structural Model}

By now, the researcher has obtained estimates of (parameters of) their model, as well as of any other objectives, including of counterfactual exercises. The researcher may wonder how to evaluate their structural model - including whether there are changes to be made, or ways to improve it. We now overview some directions. For this, it is useful to recall how such exercises are conducted in the simpler case of the linear regression model in Section (\ref{section_linear}).

In the classical linear regression model, there are four different types of validation exercises that are usually conducted. First, the authors may provide a measure of in-sample fit (e.g., $R^2$). Second, researchers may provide measures of out-of-sample fit (e.g., out-of-sample prediction, or Mean Squared Error for a sample withheld from estimation). Third, authors may test whether their model outperforms competitors (e.g., whether an alternative model is rejected in favor of their own). This may be performed by estimating nested or non-nested alternative models, and testing whether they outperform the benchmark model (e.g., by testing whether the data rejects the coefficients proposed by the alternative, or if it doesn't fit the data as well). Fourth, the authors can check whether there is some crucial component of the data that is missing: e.g., perhaps the data fails to capture tail events.

This analogy is useful because the exact same types of validation can be made in structural exercises.

First, researchers can produce measures of in-sample fit. When models are non-linear, this can include comparing the fit of moments (mean, variance etc.) of the model to observed data. 

Second, researchers can use out-of-sample measures of fit. For example, \cite{Francois22} estimate a model of factions and the organization of the Chinese Communist Party Central Committee, using individual-level politician data, including their geographical origins. They are interested in the incentives in the selection of members of the Committee but, as those choices are often secretive, they must rely on theory to bridge the gap. They use data from 1921-2012. Then, they use their model estimates to predict the composition of the 18th Party Congress in 2017. In another example, \cite{cjt} compare their model's prediction of equilibrium social connections to those observed in the data. Their estimation does not include targeted (pairwise) connections. 

This is akin to measures used in machine learning. But as many structural models are non-linear, Mean Squared Error is not an appropriate comparison statistic. Rather, one should use measures that are coherent with the model - e.g., Likelihood Ratio, when a likelihood is computable.

The third approach is to test against alternative models. A common case is to test whether the effect of interest exists (i.e., whether a certain parameter or average effect is equal to zero or not). For instance, in civil conflict, one may test whether strategic substitutability is zero or not. This can be accomplished using methods for nested models: either hypothesis testing for the relevant parameter being zero, or a Likelihood Ratio test.

In other settings, alternative models are inherently non-nested. For instance, \cite{Francois15} study cabinet allocation in African governments to test how such allocations are strategically shared across ethnicities. In the model, autocrats anticipate potential coups or revolutions when deciding how to allocate their cabinet seats. A competing theory is the ``Big Man" theory of power, where African autocrats have unconstrained decision-making ability. The latter is a theory that would generate a different cabinet allocation portfolio than the strategic sharing model of \cite{Francois15}. To test such non-nested models, the authors separately estimate each model and conduct a Vuong test (\citealp{Vuong89}). Vuong tests are particularly suited to test whether a parametric model fits better than a non-nested one, by appropriately comparing the likelihood of each model, accounting for different degrees of freedom, and checking which has the largest likelihood (i.e., it has the distribution most likely to be closer to the true one in a Kullback-Leibler sense). \cite{Shi15} provides a recent contribution that has improved properties when the models can be close to nested. 

 \section{Limits of the Structural Approach}

 It possibly goes without saying that the advantages of structural estimation clearly also come with costs.  While many have pointed out that the reduced-form approaches can struggle to tie their estimand to a theoretical quantity of interest, structural estimation avoids this issue, but at the cost of incorporating behavioral and equilibrium assumptions directly into the statistical model.  Therefore, structural estimates are really offering the reader a proposition: if you believe the model assumed is good or useful, then the estimates tell you what the real world implies about these theoretical quantities.  If you do not believe the behavioral and equilibrium assumptions, then the quantities have little real-world meaning. 

 Various assumptions, like those regarding the parametric form of random variables and the functional form of utilities, can affect the size and power of statistical tests, and possibly the identification of parameters.  Another concern is whether the behavioral model is appropriate.  In areas where formal theory is well-developed, like spatial voting or crisis bargaining, there may be little controversy.  In other areas, there may be strong theoretical disagreement or a lack of theory altogether. There is also disagreement among theorist about how seriously we should take stylized models.  In any of these cases, and many others, reduced-form empirical strategies make sense and can make contributions. Though, we emphasize that in these circumstances, reduced-form methods are still relying on different models that have their limitations. 

 It is also possible for reduced-form and structural approaches to build on synergies.  For example, a structural model could use a very credible causal effect as a moment condition to match in estimation, or as a means for model evaluation.  That is, from a full structural model, one can calculate the average treatment effect of some variable and see if it matches the one estimated by quasi-experimental means. See \cite{toddwolpin20} for a discussion about combining such approaches.

 By being clear about the link between theory and estimation, the structural approach provides significant insight into politics and informs us what political theory needs improving.  But this benefit is not costless.  Many theoretical models are very stylized and difficult to generalize, and sometimes generalized models put very few constraints on what can happen in the world.  Furthermore, data limitations can require more assumptions that may make the structural work less informative about the underlying theory, some of which may be imposed simply to make computational costs manageable.  That said, at a minimum, structural methods focus research on the conversation between theory and empirical analysis where each pushes the other  to improve our substantive understanding of the political world.

\section{Conclusion}

Empirical work should not be based on methodological camps (e.g., experimental vs. theoretical, structural vs. reduced-form), but rather based on deriving the minimal assumptions to identify and appropriately estimate the parameter of interest, whatever it may be. Occasionally, a research question may require a fully derived equilibrium model that is empirically estimated (i.e., a structural model), but sometimes it may not. Sometimes a quasi-natural experimental research design may suffice, but many times it will not. Sometimes a formal model can be embedded with an experimental design, but other times that may not be possible. It is up to us to use theory and empirical methodologies appropriately to do justice to the breadth of interesting questions in our field.

This paper overviewed structural methods, which have been increasingly used in political science. By identifying and estimating model-specific parameters, structural methods allow researchers to address a different set of questions than reduced-form analysis. This can include questions based on preferences (e.g., changes in ideological polarization), welfare (e.g., effects on equilibrium policies from changes in information), or those that require extrapolation to new environments. However, the need to carefully lay out a parameter of interest, a (statistical) model, and carefully outline identification and estimation are needed for any empirical work, whether structural or not. And this requires theory at every step of the empirical process. 

\newpage
\bibliographystyle{chicago}
\bibliography{References}

\end{document}